\documentclass
[superscriptaddress,secnumarabic,amssymb,amsmath,nobibnotes,aps,prd,showkeys,nofootinbib,nopacsnumber,onecolumn,12pt]{revtex4}%
\usepackage{graphicx}
\usepackage{bm}
\usepackage{amsmath}
\usepackage{amsfonts}
\usepackage{amssymb}%
\providecommand{\U}[1]{\protect\rule{.1in}{.1in}}

\newcommand{\be}{\begin{equation}}
\newcommand{\ee}{\end{equation}}

\newcommand{\mincir}{\raise
-3.truept\hbox{\rlap{\hbox{$\sim$}}\raise4.truept\hbox{$<$}\ }}
\newcommand{\magcir}{\raise
-3.truept\hbox{\rlap{\hbox{$\sim$}}\raise4.truept\hbox{$>$}\ }}

\begin{document}
\title{Anisotropic spacetimes in $f(T,B)$ theory I: Bianchi I Universe}
\author{Andronikos Paliathanasis}
\email{anpaliat@phys.uoa.gr}
\affiliation{Institute of Systems Science, Durban University of Technology, Durban 4000,
South Africa}
\affiliation{Instituto de Ciencias F\'{\i}sicas y Matem\'{a}ticas, Universidad Austral de
Chile, Valdivia 5090000, Chile}

\begin{abstract}
That is the first part of a series of studies on analyzing the higher-order
teleparallel theory of gravity known as the $f(T, B)$-theory. This work
attempts to understand how the anisotropic spacetimes are involved in the
$f(T, B)$-theory. In this work, we review the previous analysis of $f(T,
B)$-gravity, and we investigate the global dynamics in the case of a locally
rotational Bianchi I background geometry. We focus in the case of $f\left(
T,B\right)  =T+F\left(  B\right)  $ theory and we determine the criteria where
the $f\left(  T,B\right)  $-theory solves the homogeneity problem. Finally,
the integrability properties for the field equations are investigated by
applying the Painlev\'{e} analysis. The analytic solution is expressed by a
right Painlev\'{e} expansion.

\end{abstract}
\date{\today}
\maketitle

\section{Introduction}

\label{sec1}

The theory of General Relativity is well-tested and predicts astrophysical
objects and phenomena which observations have confirmed such as the black
holes and the gravitational waves \cite{st1,st2,st3}. Successful though
General Relativity may be in the description of astrophysical gravitational
systems, it is challenged by the analysis of the recent cosmological
observations \cite{Teg,Kowal,Komatsu,suzuki11,Ade15,ade18,hot}.

At present, the Universe is going under an acceleration phase driven by an
exotic matter source known as dark energy with negative pressure, which
provides repulsive forces and anti-gravity phenomena. Cosmological inflation
\cite{guth}, on the other hand, was introduced to solve various cosmological
problems. For example, the observable Universe's inhomogeneity and flatness.
According to the cosmic \textquotedblleft no-hair\textquotedblright%
\ conjecture \cite{nh1,nh2}, an asymptotic solution of the Universe described
by the de Sitter solution provides a rapid expansion of the size of the
Universe such that the latter effectively loses its memory of the initial
conditions, which means that the de Sitter expansion solves the
\textquotedblleft flatness\textquotedblright, \textquotedblleft
horizon\textquotedblright\ and monopole problem \cite{f1,f2}.

The modification of the Einstein-Hilbert action with the introduction of a
cosmological constant term is the most straightforward mechanism for the
description of the acceleration phases of the Universe. Indeed, a positive
cosmological constant in Bianchi cosmologies leads to expanding Bianchi
spacetimes, evolving toward the de Sitter universe \cite{w1}. However, that
conclusion is valid for the case of anisotropic background geometries such as
that of the Szekeres universes \cite{ba1,sim1}. However, the cosmological
constant suffers from other issues, for extended discussions on the
cosmological constant problems we refer the reader to \cite{rv1,rv2}.

Over the last years, cosmologists have proposed various modifications of the
Einstein-Hilbert action led to the modified theories of gravity, by
introducing geometric invariants and modifying the gravitational field
equations \cite{clifton,df1,df2}. In modified theories of gravity, the new
terms in the Einstein-Hilbert Action introduce new dynamical degrees of
freedom in the field equations which drive the dynamics such that to explain
the observational phenomena. There are various families of modified theories
of gravity, categorized according to the geometric invariants, which are
introduced in the Action Integral, see for instance
\cite{r1,r2,r3,r4,r5,r6,r7,r8,r9,r10,fg15,fg16,fg17,fg10} and references therein.

The fundamental invariant in General Relativity is the Ricciscalar $R~$of the
Levi-Civita symmetric connection. However, in the formulation of
teleparallelism and specifically in the Teleparallel Equivalent of General
Relativity (TEGR) is the fundamental geometric invariant which is used for the
definition of the gravitational Action Integral. That is, the torsion scalar
$T$ defined by the antisymmetric connection of the nonholonomic basis
\cite{ein28,Hayashi79,Maluf:1994ji,md1}, that is, of the curvature-less
Weitzenb{\"{o}}ck connection \cite{Weitzenb23}.

The equivalence of teleparallelism and General Relativity does not hold on the
modified theories of gravity based on the Ricciscalar and that based on the
torsion scalar. For instance, the $f\left(  T\right)  $-theory \cite{Ferraro}
inspired by the $f\left(  R\right)  $-theory is a second-order theory while
$f\left(  R\right)  $-theory is a fourth-order theory \cite{r1}. The latter is
because the torsion scalar $T$ admits terms with first-order derivatives,
instead of the Ricciscalar $R$, which admits second-order derivatives.
Moreover, while $f\left(  R\right)  $-gravity is equivalent to a scalar-tensor
theory, that is not true for the $f\left(  T\right)  $ theory. The two
theories are different and have different properties with that of General
Relativity \cite{ftSot,ftTam}. Applications of $f\left(  T\right)  $ theory in
the dark energy problem and in the description of the cosmological history are
presented in \cite{st1a,st2a,st3a,st4,st5,st6}.

Similarly to General Relativity and the modifications of the Einstein-Hilbert
Action Integral, there have been proposed various modified theories of gravity
based on the modification of the Action Integral for the TEGR, which extends
the $f\left(  T\right)  $ modification with the use of other invariants
constructed by the antisymmetric connection \cite{bh1}. The various families
of the modified gravitational theories based on teleparallelism are summarized
in the recent review \cite{revtel}.

We are interested in a higher-order teleparallel theory known as $f\left(
T,B\right)  $ where $B$ is the boundary term relating the torsion $T$ with the
Ricciscalar $R$, that is $B=T+R$~\cite{bh1,myr11}. Because $B$ includes
second-order derivatives, the $f\left(  T,B\right)  $-theory is a fourth-order
theory for a nonlinear function $f$ on the variable $B$. $f\left(  T,B\right)
$ provides the limits of other modified theories for specific functional forms
of $f$. Indeed, when $f\left(  T,B\right)  =f_{1}T+f_{2}B$, General Relativity
is recovered, while when $f\left(  T,B\right)  =f\left(  T\right)  +f_{2}B$,
$f\left(  T\right)  $ teleparallel theory is recovered. Finally, when
$f\left(  T,B\right)  =f\left(  T-B\right)  $, the theory gives the limit for
the $f\left(  R\right)  $-theory of gravity.

There are a plethora of studies which deal with the $f\left(  T, B\right)
$-theory in cosmology. Exact and analytic cosmological solutions with an
isotropic background space were investigated in \cite{ftb1,ftb2,ftb02}, while
the reconstruction of the cosmological history in $f\left(  T,B\right)  $
theory was the subject of study in \cite{ftb3,ftb4,ftb5,ftb6}. The presence of
nonzero spatial curvature was recently considered in \cite{ftb7}. The
minisuperspace quantization in $f\left(  T,B\right)  $ was studied in
\cite{ftb8} while an inhomogeneous exact solution was recently found in
\cite{ftb9}. The recent work \cite{ftb10} deals with the study of anisotropic
solutions in $f\left(  T,B\right)  $.\ Specifically, the Bianchi I background
space was considered, and it was found that Kasner and Kasner-like solutions
are asymptotic solutions for the field equations. However, anisotropic
exponential solutions are not preferred by the theory.

Kasner Universe \cite{kasner1} is a well-known anisotropic closed-form exact
solution of General Relativity. The Kasner four-dimensional metric has three
parameters, namely the Kasner indices, which must satisfy the two so-called
Kasner algebraic relations. There are a plethora of cosmological applications
of the Kasner solution, which makes the solution important, see, for instance
\cite{kas1,kas2,kas3,kas4,kas5,zs,bt,HT}. Anisotropic and homogeneous
cosmologies described by the Bianchi class spacetimes contain several
cosmological models that have been applied to discuss the anisotropies of the
primordial Universe and for its evolution towards the isotropy
\cite{Mis69,collins,JB1}.

This work is the first part of a series of studies where anisotropic and
homogeneous spacetimes are studied in the context of $f\left(  T, B\right)
$-theory of gravity. For the background space, we consider the locally
rotational (LRS) spacetimes, the so-called Bianchi I, Bianchi III and
Kantowski-Sachs spacetimes. These are LRS spacetimes admitting a
four-dimensional isometry group. This isometry group has a three-parameter
subgroup whose orbits are 2-surfaces of constant curvature. \cite{WE}. The
novelty of these spaces is that in the isotropic limit, the
Friedmann--Lema\^{\i}tre--Robertson--Walker (FLRW) Universe is recovered.
Indeed, the Bianchi I reduces to the spatially flat FLRW spacetime, while the
open and closed FLRW spacetimes follow from the isotropic Bianchi III and
Kantowski-Sachs spacetimes, respectively. The Kantowski-Sachs spacetime can
follow from the LRS Bianchi type IX metric by a Lie contraction \cite{WE}.
Additionally, these spacetimes can be seen as the limit of the homogeneity in
the case of the silent Universe and specifically of the Szekeres spacetimes
\cite{szek0}.

We perform a detailed analysis of the dynamics of the field equations by using
normal coordinates to investigate their asymptotic behaviour. Indeed, we shall
study the existence of exact anisotropic solutions. We also construct the
cosmological history by assuming an anisotropic background space, with or
without spatial curvature. Such an analysis will provide us with necessary
information about the $f\left(  T,B\right)  $-theory. We shall investigate if
isotropy and flatness problems can be solved by the $f\left(  T,B\right)
$-theory, as well as if the limit of General Relativity can be recovered. This
approach has been widely studied in various gravitational theories with many
exciting results, for instance
\cite{cop1,mo1,mo2,mo3,mo5,mo6,mo8,mo9,cher1,and3,mmf1,mmf4,col112,col113,col11,ae4,ans1,dn1}%
.

In this study, we present a brief review of the $f\left(  T,B\right)  $-theory
and we focus on the $f\left(  T,B\right)  =T+F\left(  B\right)  $ theory for
the anisotropic and homogeneous Bianchi I geometry. The $f\left(  T,B\right)
=T+F\left(  B\right)  $ theory is of special interest because it is a
fourth-order theory of gravity with the same degrees of freedom as the
$f\left(  R\right)  $-theory. Moreover, there exists a scalar field
description related to scalar-torsion theory. Anisotropic and homogeneous
background geometries are investigated in the studies \cite{paper2}, and
\cite{paper3}. Indeed, the Kantowski-Sachs spacetime and the LRS Bianchi III
spacetimes are considered, respectively. As we have mentioned before there is
a one-to-one relation of the anisotropic geometries of our consideration with
the FLRW spacetimes in the isotropization limit. Because the three different
geometries provide different topological spaces in terms of the dynamical
systems we decided to split this piece of study in a series of works. In
addition, substantial differences are expected in the cosmological evolution
of the physical variables according to whether the spacelike surface has open,
closed or flat topology, which make necessary the need to separate the present
analysis in a series of studies according to the admitted topology, see also
the discussion in \cite{paper2}.

Finally, in \cite{paper4}, we focus on the existence of the minisuperspace
description of the field equations. We apply the Noether symmetry analysis
\cite{ns1} to construct conservation laws and determine exact and analytic
solutions for the field equations. The plan of the paper is as follows.

In Section \ref{sec2} we present a brief review of the teleparallel $f\left(
T,B\right)  $-theory where we discuss the difference between the TEGR, the
$f\left(  T\right)  $ and the $f\left(  T,B\right)  $ theories. For the
Bianchi I geometry, in Section \ref{sec3} we derive the field equations and
explain how the function form of the $f\left(  T, B\right)  $ theory affects
the geometrodynamical degrees of freedom which are involved in the
cosmological dynamics. The dynamical system analysis of the $f\left(
T,B\right)  =T+F\left(  B\right)  $ theory is presented in Section \ref{sec4}.
We apply the $H$-normalization approach and write the field equations in an
equivalent form using dimensionless variables. For the new system, we
determine the stationary points. Each stationary point corresponds to a
specific asymptotic solution. The physical properties of the asymptotic
solutions are investigated, along with the stability properties. They are an
essential analysis in order to reconstruct the cosmological history and answer
the question if the $f\left(  T, B\right)  =T+F\left(  B\right)  $ can solve
the isotropy of the Universe for anisotropic initial conditions. In Section
\ref{sec5}, we investigate the integrability properties of the field equations
and the existence of an analytic solution by using the Painlev\'{e} analysis.
Finally, Section \ref{sec6} summarises the results and conclusions.

\section{Teleparallel theory of gravity}

\label{sec2}

In teleparallelism the vierbein fields ${\mathbf{e}_{\mu}(x^{\sigma})}$
introduce the dynamical degrees of freedom \cite{Hayashi79}. They form an
orthonormal basis for the tangent space at each point $P$ such that $g(e_{\mu
},e_{\nu})=\mathbf{e}_{\mu}\cdot\mathbf{e}_{\nu}=\eta_{\mu\nu}$, where
$\eta_{\mu\nu}~$is the line element of the Minkowski spacetime, $\eta_{\mu\nu
}=\text{diag}\left(  -,+,+,+\right)  $. The commutator relations for the
vierbein fields are $\ [e_{\mu},e_{\nu}]=c_{\nu\mu}^{\beta}e_{\beta}~$\ where
$c_{\left(  \nu\mu\right)  }^{\beta}=0.$

In the nonholonomic coordinates the covariant derivative $\nabla_{\mu}$ is
defined with the connection
\begin{equation}
\mathring{\Gamma}_{\nu\beta}^{\mu}=\{_{\nu\beta}^{\mu}\}+\frac{1}{2}%
g^{\mu\sigma}(c_{\nu\sigma,\beta}+c_{\sigma\beta,\nu}-c_{\mu\beta,\sigma}),
\label{cc.02}%
\end{equation}
where $\{_{\nu\beta}^{\mu}\}$ is the symmetric Levi-Civita connection of
Riemannian geometry which is used in General Relativity.

For the case where $\mathbf{e}_{\mu}\cdot\mathbf{e}_{\nu}=\eta_{\mu\nu}$,~the
Weitzenb{\"{o}}ck connection \cite{Weitzenb23}%
\begin{equation}
\mathring{\Gamma}_{\nu\beta}^{\mu}=\frac{1}{2}\eta^{\mu\sigma}(c_{\nu
\sigma,\beta}+c_{\sigma\beta,\nu}-c_{\mu\beta,\sigma}), \label{cc.03}%
\end{equation}
where $\mathring{\Gamma}_{\nu\beta}^{\mu}$ are antisymmetric in the two first
indices, that is
\begin{equation}
\mathring{\Gamma}_{\mu\nu\beta}=-\mathring{\Gamma}_{\nu\mu\beta},
~\mathring{\Gamma}_{\mu\nu\beta}=\eta_{\mu\sigma}\mathring{\Gamma}_{\nu\beta
}^{\mu},
\end{equation}
and they describe the Ricci rotation coefficients.

The nonnull torsion tensor is defined by the relation
\begin{equation}
T_{\mu\nu}^{\beta}=\mathring{\Gamma}_{\nu\mu}^{\beta}-\mathring{\Gamma}%
_{\mu\nu}^{\beta},
\end{equation}
while the torsion scalar $T$ is given by the expression$\ $%
\begin{equation}
T=\frac{1}{2}({K^{\mu\nu}}_{\beta}+\delta_{\beta}^{\mu}{T^{\theta\nu}}%
_{\theta}-\delta_{\beta}^{\nu}{T^{\theta\mu}}_{\theta}){T^{\beta}}_{\mu\nu}.
\end{equation}
where
\begin{equation}
K_{~~~\beta}^{\mu\nu}=-\frac{1}{2}({T^{\mu\nu}}_{\beta}-{T^{\nu\mu}}_{\beta
}-{T_{\beta}}^{\mu\nu}).
\end{equation}

In TEGR the gravitational Action Integral is as follows \cite{ein28}
\begin{equation}
S_{T}=\frac{1}{16\pi G}\int d^{4}xeT~+S_{m},~e=\det(e_{\mu}), \label{cc.05}%
\end{equation}
in which $S_{m}$ is the Action Integral component for the matter source.

\subsection{$f\left(  T\right)  $-theory}

The simplest extension of TEGR is the $f\left(  T\right)  $. The gravitational
Action Integral (\ref{cc.05}) is modified as follow \cite{Ferraro}
\begin{equation}
S_{f\left(  T\right)  }=\frac{1}{16\pi G}\int d^{4}xe\left(  f(T)\right)
+S_{m}, \label{ft.05}%
\end{equation}
where $f$ is a function which should be considered. In the limit where
$f\left(  T\right)  $ is a linear function TEGR is recovered. Because $T$
admits only first order derivatives of the vierbeins, then the gravitational
field equations are of second-order.

Variation with respect to the vierbein provides the modified field equations
\cite{revtel}%
\begin{equation}
4\pi Ge\mathcal{T}_{a}^{\left(  m\right)  }{}^{\lambda}=ef_{,T}G_{a}^{\lambda
}+\left[  \frac{1}{4}\left(  Tf_{,T}-f\right)  eh_{a}^{\lambda}+e(f_{,T}%
)_{,\mu}S_{a}{}^{\mu\lambda}\right]  , \label{ft.0}%
\end{equation}
with $\mathcal{T}_{a}^{\left(  m\right)  }{}^{\lambda}$ the energy--momentum
tensor of the matter source~$S_{m}$ while $f_{T}$ and $f_{TT}$ denote the
first and second derivatives of the function $f(T)$ with respect to $T$,
$S_{a}{}^{\mu\lambda}$ is the superpotential tensor
\begin{equation}
{S_{\beta}}^{\mu\nu}=\frac{1}{2}({K^{\mu\nu}}_{\beta}+\delta_{\beta}^{\mu
}{T^{\theta\nu}}_{\theta}-\delta_{\beta}^{\nu}{T^{\theta\mu}}_{\theta}),
\label{ft.03}%
\end{equation}
and $e_{\alpha}=h_{\alpha}^{\lambda}\left(  x\right)  \partial_{\lambda}$ is
the coordinate basis for the vierbein fields.

From (\ref{ft.0}) it follows that the limit of TEGR is recovered for a linear
Lagrangian function $f_{,TT}=0$. However, that is not the only possible case.
Vacuum solutions of General Relativity follow in the case of a nonlinear
function $f\left(  T\right)  $ when \cite{st6}
\begin{equation}
T=0, ~f\left(  T\right)  _{|T\rightarrow0}=0,
\end{equation}
and%
\begin{equation}
S_{\rho}{}^{\mu\nu}\partial_{\mu}({T})f_{TT}=0.
\end{equation}

Similarly, solutions with cosmological constant term follow when \cite{rfa}%
\begin{equation}
f\left(  T\right)  _{|T\rightarrow-\Lambda}=0, ~T=-\Lambda\text{.}%
\end{equation}

\subsection{$f\left(  T,B\right)  $-theory}

The $f\left(  T,B\right)  $-theory is a natural extension of $f\left(
T\right)  $ teleparallel theory where the boundary term
\begin{equation}
B=2e^{-1}\partial_{\nu}\left(  eT_{\rho}^{~\rho\nu}\right)  ,
\end{equation}
$~$which connects the torsion scalar $T$ with the Ricciscalar $R$, i.e.
\begin{equation}
B=T+R,
\end{equation}
is used in the gravitational Action Integral.

The modified gravitational Action Integral is \cite{bh1}%
\begin{equation}
S_{f\left(  T,B\right)  }=\frac{1}{16\pi G}\int d^{4}xef\left(  T,B\right)  .
\label{cc.06}%
\end{equation}

Variation for the vierbein fields provides the field equations%

\begin{align}
4\pi Ge\mathcal{T}_{a}^{\left(  m\right)  }{}^{\lambda}  &  =ef_{,T}%
G_{a}^{\lambda}+\left[  \frac{1}{4}\left(  Tf_{,T}-f\right)  eh_{a}^{\lambda
}+e(f_{,T})_{,\mu}S_{a}{}^{\mu\lambda}\right] \nonumber\\
&  +\left[  e(f_{,B})_{,\mu}S_{a}{}^{\mu\lambda}-\frac{1}{2}e\left(
h_{a}^{\sigma}\left(  f_{,B}\right)  _{;\sigma}^{~~~;\lambda}-h_{a}^{\lambda
}\left(  f_{,B}\right)  ^{;\mu\nu}g_{\mu\nu}\right)  +\frac{1}{4}%
eBh_{a}^{\lambda}f_{,B}\right]  . \label{cc.08}%
\end{align}

Where the terms
\begin{equation}
T_{a}^{\left(  B\right)  \lambda}=e(f_{,B})_{,\mu}S_{a}{}^{\mu\lambda}%
-\frac{1}{2}e\left(  h_{a}^{\sigma}\left(  f_{,B}\right)  _{;\sigma
}^{~~~;\lambda}-h_{a}^{\lambda}\left(  f_{,B}\right)  ^{;\mu\nu}g_{\mu\nu
}\right)  +\frac{1}{4}eBh_{a}^{\lambda}f_{,B},
\end{equation}
are related to a nonlinear function $f$ on the variable $B$, which includes
the higher-order derivatives.

Similarly to before the limit of TEGR is recovered when $f\left(  T,B\right)
$ is a linear function, that is, $f\left(  T,B\right)  =f_{1}T+f_{2}B$. The
special case $f\left(  T,B\right)  =T+F\left(  B\right)  $ has drawn the
attention before because for small values of $F\left(  B\right)  $ there
exists a small derivation from TEGR.

In addition the additional degrees of freedom of the field equations can be
attributed to a scalar field. Thus is for $\phi=f_{,B}$, terms $T_{a}^{\left(
B\right)  \lambda}$ become%
\begin{equation}
T_{a}^{\left(  B\right)  \lambda}=e\phi_{,\mu}S_{a}{}^{\mu\lambda}-\frac{1}%
{2}e\left(  h_{a}^{\sigma}\phi_{;\sigma}^{~~~;\lambda}-h_{a}^{\lambda}%
\phi^{;\mu\nu}g_{\mu\nu}\right)  -\frac{1}{4}eh_{a}^{\lambda}V\left(
\phi\right)  +\frac{1}{4}eh_{a}^{\lambda}f,
\end{equation}
with $V\left(  \phi\right)  =f-Bf_{,B}$. At this point it is important to
mention that $f\left(  T,B\right)  $ gravity under a conformal transformation
is equivalent to the scalar-torsion theory \cite{dm1,dm2}. That is analogue to
the relation of $f\left(  R\right)  $-gravity with the scalar field theories.
$f\left(  T,B\right)  $ theory and $f\left(  R\right)  $-gravity are of the
same order, while the $f\left(  T\right)  =T+F\left(  B\right)  $ theory has
the same degrees of freedom as the $f\left(  R\right)  $-theory and canonical
quantization in the minisuperspace approach can be applied \cite{ftb8}.

\section{Bianchi I Universe}

\label{sec3}

In the Misner variables the LRS Bianchi I spacetime is described by the line
element
\begin{equation}
ds^{2}=-N^{2}\left(  t\right)  dt^{2}+e^{2\alpha\left(  t\right)  }\left(
e^{2\beta\left(  t\right)  }dx^{2}+e^{-\beta\left(  t\right)  }\left(
dy^{2}+dz^{2}\right)  \right)  , \label{ch.03}%
\end{equation}
where $\alpha\left(  t\right)  $ is the scale factor for the three-dimensional
hypersurface and $\beta$ is the anisotropic parameter while $N\left(
t\right)  $ is the lapse function. For $\beta\left(  t\right)  \rightarrow0$,
the line element (\ref{ch.03}) reduces to the spatially flat FLRW geometry.

In order to calculate the torsion scalar $T$, we should consider the vierbein
fields \cite{revtel}, thus we assume the basis
\[
e^{1}=Ndt, ~e^{2}=e^{\alpha+\beta}dx, ~e^{3}=e^{\alpha-\frac{\beta}{2}}dy,
~e^{4}=e^{\alpha-\frac{\beta}{2}}dz.
\]
which provides%
\begin{equation}
T=\frac{1}{N^{2}}\left(  6\dot{\alpha}^{2}-\frac{3}{2}\dot{\beta}^{2}\right)
, \label{ch.04}%
\end{equation}
and the boundary term%
\begin{equation}
B=\frac{6}{N^{2}}\left(  \ddot{\alpha}-\dot{\alpha}\frac{\dot{N}}{N}%
+3\dot{\alpha}^{2}\right)  , \label{ch.05}%
\end{equation}
while the Lagrangian of General Relativity, i.e. the Ricciscalar, is
calculated
\begin{equation}
R=\frac{1}{N^{2}}\left(  6\ddot{\alpha}-6\dot{\alpha}\frac{\dot{N}}{N}%
+12\dot{\alpha}^{2}+\frac{3}{2}\dot{\beta}^{2}\right)  . \label{ch.06}%
\end{equation}

Bianchi I spacetime admits a minisuperspace description, which means that a
point-like Lagrangian function reproduces the existing field equations. In
order to write the point-like Lagrangian, we shall use the method of Lagrange
multiplier. The latter approach is a powerful method widely applied in various
modified theories of gravity; see, for instance, \cite{lan1,lan2,lan3} and
references therein.

In the case of vacuum\ the Action Integral (\ref{cc.06}) with the use of two
Lagrange multipliers $\lambda_{1}$ and $\lambda_{2}$ for the Bianchi I
background space can be written in the equivalent form%
\begin{align}
S_{f\left(  T,B\right)  }  &  =\frac{1}{16\pi G}\int d^{4}xNe^{3\alpha}\left(
f(T,B)\right) \nonumber\\
&  +\frac{1}{16\pi G}\int d^{4}xNe^{3\alpha}\lambda_{1}\left(  T-\frac
{1}{N^{2}}\left(  6\dot{\alpha}^{2}-\frac{3}{2}\dot{\beta}^{2}\right)  \right)
\nonumber\\
&  +\frac{1}{16\pi G}\int d^{4}xNe^{3\alpha}\lambda_{2}\left(  B-\frac
{6}{N^{2}}\left(  \ddot{\alpha}-\dot{\alpha}\frac{\dot{N}}{N}+3\dot{\alpha
}^{2}\right)  \right)  . \label{ch.07}%
\end{align}

Variation with respect to the variables $T$ and $B$, it follows $\frac{\delta
}{\delta T}S_{f\left(  T,B\right)  }=0$, $\frac{\delta}{\delta B}S_{f\left(
T,B\right)  }=0$, that is,%
\begin{equation}
\lambda_{1}=-f_{,T}\text{ and }\lambda_{2}=-f_{,B}. \label{ch.08}%
\end{equation}

Thus, by replacing in (\ref{ch.07}) and integrating by parts we end with the
point-like Lagrangian%

\begin{equation}
\mathcal{L}\left(  \alpha,\dot{\alpha},\beta,\dot{\beta},T,B,\dot{B}\right)
=\frac{1}{N}\left(  f_{,T}e^{3\alpha}\left(  6\dot{\alpha}^{2}-\frac{3}{2}%
\dot{\beta}^{2}\right)  -6e^{3\alpha}f_{,BB}\dot{\alpha}\dot{B}\right)
+Ne^{3\alpha}\left(  f-Tf_{,T}-Bf_{,B}\right)  . \label{ch.09}%
\end{equation}
\qquad

In the following analysis we focus on the case where $f\left(  T,B\right)  $
is separable, that is,
\begin{equation}
f\left(  T,B\right)  =T+K\left(  T\right)  +F\left(  B\right)  \text{, }%
\end{equation}
such that Lagrangian (\ref{ch.08}) to become%
\begin{align}
\mathcal{L}\left(  \alpha,\dot{\alpha},\beta,\dot{\beta},T,B,\dot{B}\right)
&  =\frac{1}{N}\left(  \left(  1+K_{,T}\right)  e^{3\alpha}\left(
6\dot{\alpha}^{2}-\frac{3}{2}\dot{\beta}^{2}\right)  -6e^{3\alpha}F_{,BB}%
\dot{\alpha}\dot{B}\right) \nonumber\\
&  +Ne^{3\alpha}\left(  K-TK_{,T}+F\left(  B\right)  -BF_{, B}\right)  ,
\end{align}
or equivalently%
\begin{align}
\mathcal{L}\left(  \alpha,\dot{\alpha},\beta,\dot{\beta},T,\phi,\dot{\phi
}\right)   &  =\frac{1}{N}\left(  \left(  1+K_{,T}\right)  e^{3\alpha}\left(
6\dot{\alpha}^{2}-\frac{3}{2}\dot{\beta}^{2}\right)  -6e^{3\alpha}\dot{\alpha
}\dot{\phi}\right) \nonumber\\
&  +Ne^{3\alpha}\left(  K-TK_{,T}+V\left(  \phi\right)  \right)  ,
\label{ch.12}%
\end{align}
with $\phi=F_{,B}$ and $V\left(  \phi\right)  =F\left(  B\right)  -BF_{, B}$.

Thus, the gravitational field equations are
\begin{equation}
0=\left(  \left(  1+K_{,T}\right)  e^{3\alpha}\left(  6\dot{\alpha}^{2}%
-\frac{3}{2}\dot{\beta}^{2}\right)  -6e^{3\alpha}\dot{\alpha}\dot{\phi
}\right)  -e^{3\alpha}\left(  K-TK_{,T}+V\left(  \phi\right)  \right)  ,
\label{ee.01}%
\end{equation}%
\begin{equation}
0=\left(  1+K_{,T}\right)  \left(  \ddot{\alpha}+\frac{3}{2}\dot{\alpha}%
^{2}+\frac{3}{8}\dot{\beta}^{2}\right)  -\frac{1}{4}\left(  K-TK_{,T}\right)
+K_{,TT}\dot{\alpha}\dot{T}-\frac{1}{2}\left(  \ddot{\phi}+\frac{1}{2}V\left(
\phi\right)  \right)  , \label{ee.02}%
\end{equation}
\begin{equation}
0=\ddot{\beta}+\dot{\beta}\left(  3\dot{\alpha}+\dot{T}K_{,TT}\right)  ,
\label{ee.03}%
\end{equation}
and the two constraint equations%
\begin{equation}
0=V_{,\phi}+6\left(  3\dot{\alpha}^{2}+\ddot{\alpha}\right)  , \label{ee.04}%
\end{equation}%
\begin{equation}
0=\left(  6\dot{\alpha}^{2}-\frac{3}{2}\dot{\beta}^{2}\right)  -T,
\label{ee.05}%
\end{equation}
where without loss of generality we have selected the lapse function $N\left(
t\right)  $ to be constant, that is, $N\left(  t\right)  =1$.

We define the Hubble variable $H=\dot{a}$ and we write equations
(\ref{ee.01}), (\ref{ee.02}) in the equivalent form%
\begin{equation}
\left(  3H^{2}-\frac{3}{4}\dot{\beta}^{2}\right)  =G_{eff}\rho_{f\left(
T,B\right)  },
\end{equation}%
\begin{equation}
\left(  2\dot{H}+3H^{2}-\frac{3}{4}\dot{\beta}^{2}\right)  =-G_{eff}%
p_{f\left(  T,B\right)  },
\end{equation}
in which the geometrodynamical fluid source has the following components%
\begin{equation}
\rho_{f\left(  T,B\right)  }=\left(  6H\dot{\phi}+\frac{1}{2}\left(
K-TK_{,T}+V\left(  \phi\right)  \right)  \right)  ,
\end{equation}%
\begin{equation}
p_{f\left(  T,B\right)  }=-\frac{1}{2}\left(  K-TK_{,T}\right)  +2K_{,TT}%
H\dot{T}+\left(  \ddot{\phi}+\frac{1}{2}V\left(  \phi\right)  \right)  ,
\end{equation}
and \ $G_{eff}=\left(  1+K_{,T}\right)  ^{-1}$ is the effective gravitational
constant, which varies on time for a nonlinear function $K\left(  T\right)  $.

However that is not the unique way to define the geometrodynamical fluid.
Indeed, without loss of generality we can write the field equations
\begin{equation}
\left(  3H^{2}-\frac{3}{4}\dot{\beta}^{2}\right)  =\bar{\rho}_{f\left(
T,B\right)  },
\end{equation}%
\[
\left(  2\dot{H}+3H^{2}+\frac{3}{4}\dot{\beta}^{2}\right)  =-\bar{p}_{f\left(
T,B\right)  },
\]
where now%
\begin{equation}
\bar{\rho}_{f\left(  T,B\right)  }=\left(  3H\dot{\phi}-\frac{1}{2}\left(
K-2TK_{,T}+V\left(  \phi\right)  \right)  \right)  ,
\end{equation}%
\begin{align}
\bar{p}_{f\left(  T,B\right)  }  &  =-\frac{1}{2}\left(  K-TK_{,T}\right)
+2K_{,TT}H\dot{T}\nonumber\\
&  +\left(  \ddot{\phi}+\frac{1}{2}V\left(  \phi\right)  \right)
+K_{,T}\left(  2\dot{H}+3H^{2}+\frac{3}{4}\dot{\beta}^{2}\right)  .
\end{align}

In the special case of $f\left(  T,B\right)  =T+F\left(  B\right)  ,~$where
$G_{eff}=cont$, the modified field equations read%
\begin{equation}
\left(  3H^{2}-\frac{3}{4}\dot{\beta}^{2}\right)  =\left(  3H\dot{\phi}%
+\frac{1}{2}V\left(  \phi\right)  \right)  ,
\end{equation}%
\begin{equation}
\left(  2\dot{H}+3H^{2}+\frac{3}{4}\dot{\beta}^{2}\right)  =-\left(
\ddot{\phi}+\frac{1}{2}V\left(  \phi\right)  \right)  ,
\end{equation}%
\begin{equation}
0=\ddot{\beta}+3\dot{\beta}\dot{\alpha},
\end{equation}
where now the geometrodynamical terms follow only from the boundary term
$F\left(  B\right)  $.

Furthermore, for $\phi=const$, the $f\left(  T\right)  $-theory is recovered
and the field equations are
\begin{equation}
\left(  3H^{2}-\frac{3}{4}\dot{\beta}^{2}\right)  =\frac{\frac{1}{2}\left(
K-TK_{,T}\right)  }{1+K_{,T}},
\end{equation}%
\begin{equation}
\left(  2\dot{H}+3H^{2}+\frac{3}{4}\dot{\beta}^{2}\right)  =-\frac{-\frac
{1}{2}\left(  K-TK_{,T}\right)  +2K_{,TT}H\dot{T}}{1+K_{,T}},
\end{equation}
or equivalently%
\begin{equation}
\left(  3H^{2}-\frac{3}{4}\dot{\beta}^{2}\right)  =\frac{1}{2}\left(
K-2TK_{,T}\right)  ,
\end{equation}%
\begin{align}
-\left(  2\dot{H}+3H^{2}+\frac{3}{4}\dot{\beta}^{2}\right)   &  =-\frac{1}%
{2}\left(  K-TK_{,T}\right) \nonumber\\
&  +2K_{,TT}H\dot{T}+K_{,T}\left(  2\dot{H}+3H^{2}+\frac{3}{4}\dot{\beta}%
^{2}\right)  ,
\end{align}
while for the anisotropic parameter it holds%
\begin{equation}
0=\ddot{\beta}+\dot{\beta}\left(  3\dot{\alpha}+\dot{T}K_{,TT}\right)  .
\end{equation}

\section{Dynamical analysis for $f\left(  T,B\right)  =T+F\left(  B\right)  $
gravity}

\label{sec4}

We want to extend the analysis presented in \cite{ftb1,ftb3} for the case of
anisotropic spacetimes. That is, we focus on the case where $f\left(
T,B\right)  =T+F\left(  B\right)  $. This specific theory has many interesting
properties. It is a modified teleparallel theory with many degrees of freedom
as the $f\left(  R\right)  $-theory, and it admits a scalar field description.
From the previous analysis, it is clear that this specific theory can describe
interesting epochs of cosmological history in the case of isotropy.

\begin{figure}[th]
\centering\includegraphics[width=1\textwidth]{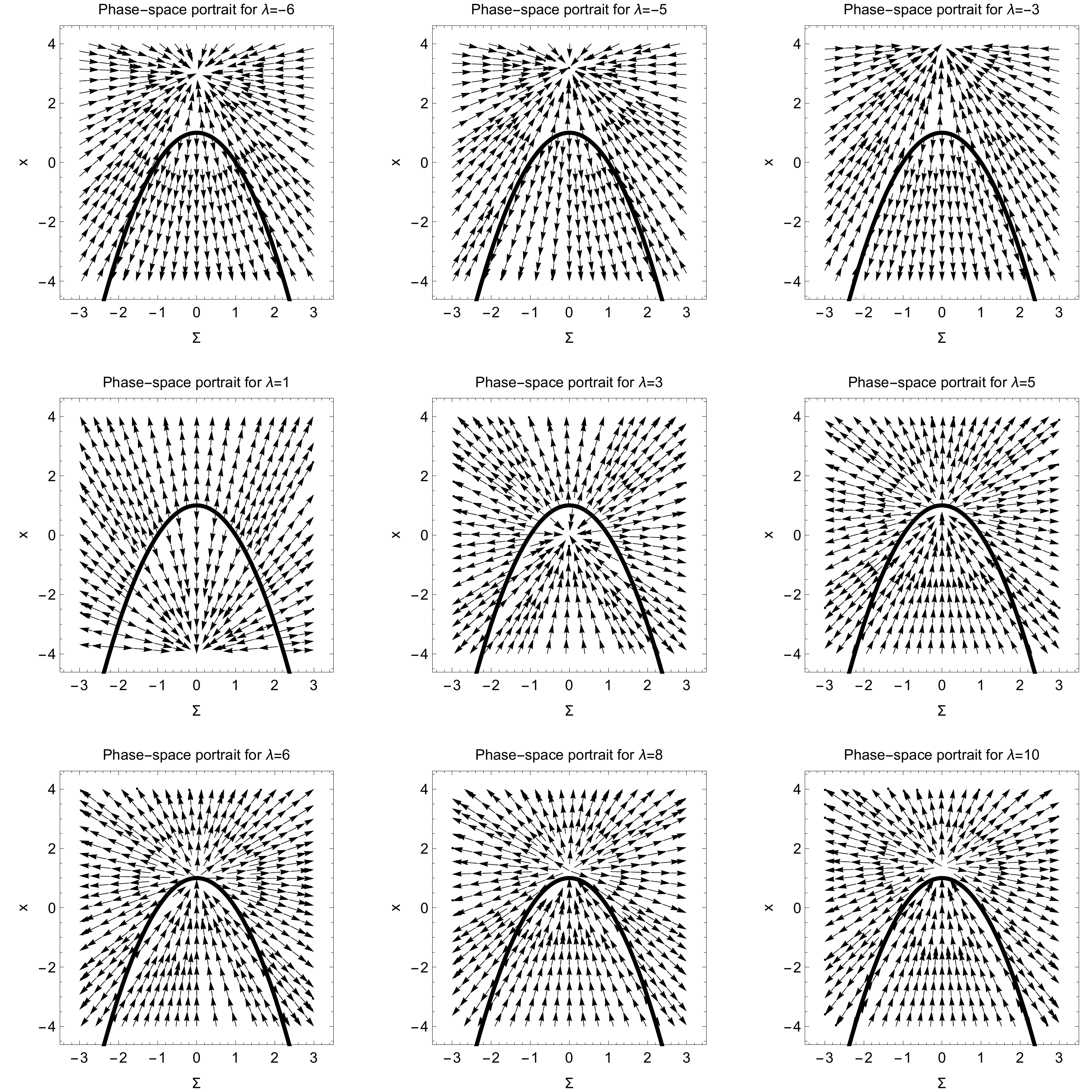}\caption{Phase-space
portrait for the two-dimensionless dynamical system (\ref{sd.06}),
(\ref{sd.07}) in the finite regime for different values of the free parameter
$\lambda$. The Figs. of the first and second row are for $\lambda<6$ where
$P_{2}$ is the unique attractor. However for $\lambda\geq6$ the dynamical
system goes at infinity and $P_{2}$ is a source point. The solid line
corresponds to the family of points $P_{1}$ where it is clear that the
asymptotic solutions at the points $P_{1}~$are always unstable. }%
\label{bf1}%
\end{figure}

In order to perform a detailed analysis of the dynamics we define new
dimensionless variables%
\begin{equation}
\Sigma=\frac{\dot{\beta}}{2H}, ~x=\frac{\dot{\phi}}{H}, ~y=\frac{V\left(
\phi\right)  }{6H^{2}}, ~\lambda=-\ln\left(  V\left(  \phi\right)  \right)  .
\label{sd.01}%
\end{equation}

Moreover, we define the new independent variable $d\tau=Hdt$ and the
exponential potential $V\left(  \phi\right)  =V_{0}e^{-\lambda\phi}$ such that
the field equations to be%
\begin{equation}
\frac{d\Sigma}{d\tau}=-\lambda y\Sigma, \label{sd.02}%
\end{equation}%
\begin{equation}
\frac{dx}{d\tau}=\left(  2\lambda-3\right)  y+x\left(  3-\lambda y\right)
+3\left(  \Sigma^{2}-1\right)  , \label{sd.03}%
\end{equation}%
\begin{equation}
\frac{dy}{d\tau}=-y\left(  \lambda\left(  x+2y\right)  -6\right)  ,
\label{sd.04}%
\end{equation}
with constraint equation
\begin{equation}
1-x-y-\Sigma^{2}=0. \label{sd.05}%
\end{equation}

For the exponential potential we find that the $F\left(  B\right)  $ function
is $F\left(  B\right)  =-\frac{1}{\lambda}B\ln B~$\cite{ftb1}.

With the use of the constraint equation (\ref{sd.05}) we replace variable
$y=1-x-\Sigma^{2}$ in the dynamical system and we end with the set of
differential equations
\begin{equation}
\frac{d\Sigma}{d\tau}=-\lambda\left(  1-x-\Sigma^{2}\right)  \Sigma,
\label{sd.06}%
\end{equation}%
\begin{equation}
\frac{dx}{d\tau}=\left(  2\lambda-3\right)  \left(  1-x-\Sigma^{2}\right)
+x\left(  3-\lambda\left(  1-x-\Sigma^{2}\right)  \right)  +3\left(
\Sigma^{2}-1\right)  , \label{sd.07}%
\end{equation}

Variables $\left(  \Sigma,x\right)  $ are not constrained, thus, they can take
values at all the region of the real values, i.e. $\left(  x,\Sigma\right)
\in%
\mathbb{R}
$. Consequently, we shall study the existence of stationary points for the
two-dimensional dynamical system (\ref{sd.06}) and (\ref{sd.07}) in the finite
and infinity regimes.

Each stationary point of the dynamical system corresponds to a specific era of
cosmological evolution. The dependent variables' values determine the
asymptotic solution's physical properties. For each stationary point, we
investigate the stability properties. The latter is a fundamental analysis to
construct the cosmological evolution and determine the future attractors.

\subsection{Analysis at the finite regime}

The stationary points~$P\left(  \Sigma\left(  P\right)  ,x\left(  P\right)
\right)  $ for the dynamical system (\ref{sd.06}), (\ref{sd.07}) are%
\begin{equation}
P_{1}=\left(  \Sigma,1-\Sigma^{2}\right)  , ~P_{2}=\left(  0,2-\frac
{6}{\lambda}\right)  . \label{sd.08}%
\end{equation}

Points $P_{1}$ describe a family of anisotropic spacetimes where there is not
any contribution of the potential term to the cosmological solution, that is,
$y\left(  P_{1}\right)  =0$. The eigenvalues of the linearized system around
$P_{1}$ are $e_{1}\left(  P_{1}\right)  =0$ and $e_{2}\left(  P_{1}\right)
=\lambda\left(  \Sigma^{2}-1\right)  +6$. For $e_{2}\left(  P_{1}\right)  >0$
the stationary point is always a source, however, for $e_{2}\left(
P_{1}\right)  <0$ the center manifold theorem should be applied because may
there exist a stable submanifold. However, the application of the center
manifold theorem does not contribute in the physical discussion of the theory,
such that we omit it.

Point $P_{2}$ corresponds to an asymptotic solution which describes an
isotropic spatially flat FLRW Universe where the cosmological fluid has the
equation of state parameter $w_{eff}\left(  P_{2}\right)  =-3+\frac{2}%
{3}\lambda.$ Thus for $\lambda=3$ the de Sitter universe is recovered. The
eigenvalues of the linearized system are $e_{1}\left(  P_{2}\right)
=\lambda-6$ and $e_{2}\left(  P_{2}\right)  =\lambda-6$, thus for $\lambda<6$
the stationary point is an attractor.

In Fig. \ref{bf1} we present the two-dimensional phase-space portrait for the
dynamical system (\ref{sd.06}), (\ref{sd.07}) in the finite regime. We observe
that points $P_{1}$ always \ describe unstable anisotropic spacetimes.
Furthermore, we conclude that for $\lambda<6$, the cosmological model has the
homogeneous and isotropic spatially flat Universe as an attractor.

The results are summarized in Table \ref{tab1}.%

\begin{table}[tbp] \centering
\caption{Stationary points at the finite regime}%
\begin{tabular}
[c]{ccccc}\hline\hline
\textbf{Point} & \textbf{Existence} & \textbf{Spacetime} & $\mathbf{q<0}$ &
\textbf{Stable?}\\\hline
$P_{1}$ & Always & Bianchi I & No & No\\
$P_{2}$ & $\lambda\neq0$ & FLRW (Flat) & $\lambda<4$ & $\lambda<6$%
\\\hline\hline
\end{tabular}
\label{tab1}%
\end{table}%

\subsection{Analysis at the infinity}

In order to understand the global dynamics of the dynamic system
(\ref{sd.06}), (\ref{sd.07}) we investigate the existence of stationary points
at the infinity.

\begin{figure}[th]
\centering\includegraphics[width=1\textwidth]{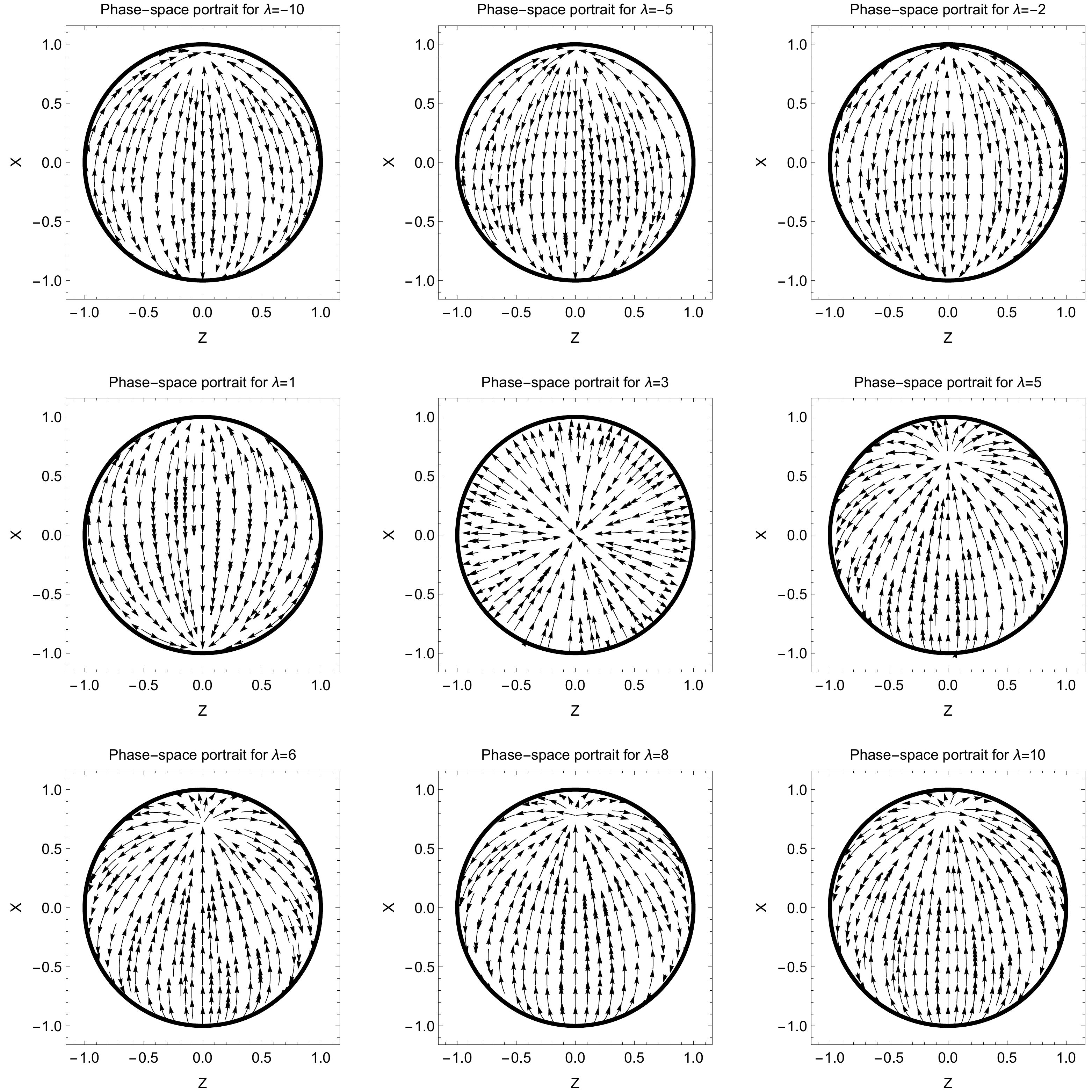}\caption{Phase-space
portrait for the two-dimensionless dynamical system (\ref{sd.09}),
(\ref{sd.10}) for different values of the free parameter $\lambda$. The Figs.
of the first and second row are for $\lambda<6$ where $P_{2}$ is the unique
attractor. However for $\lambda\geq6$ the dynamical system goes at infinity
and $P_{2}$ is a source point. }%
\label{bf2}%
\end{figure}

We define the new Poincar\'{e} variables%
\begin{equation}
\Sigma=\frac{Z}{\sqrt{1-X^{2}-Z^{2}}}, ~x=\frac{X}{\sqrt{1-X^{2}-Z^{2}}},
~d\eta=\sqrt{1-X^{2}-Z^{2}}d\tau. \label{sd.09}%
\end{equation}

Thus, the dynamical system (\ref{sd.06}), (\ref{sd.07}) reads%
\begin{align}
\frac{1}{Z}\frac{dZ}{d\eta}  &  =X^{3}\left(  \lambda-6\right)  -X\left(
\lambda-6-3\left(  \lambda-4\right)  Z^{2}\right) \label{sd.10}\\
&  +\sqrt{1-X^{2}-Z^{2}}\left(  3\left(  \lambda-2\right)  X^{2}%
+\lambda\left(  2Z^{2}-1\right)  \right)  ,\nonumber
\end{align}%
\begin{align}
\frac{dX}{d\eta}  &  =X^{4}\left(  \lambda-6\right)  +3X^{2}\left(
\lambda-4\right)  \left(  Z^{2}-1\right)  -2\left(  \lambda-3\right)  \left(
2Z^{2}-1\right) \nonumber\\
&  +\sqrt{1-X^{2}-Z^{2}}\left(  3\left(  \lambda-2\right)  \left(
X^{2}-1\right)  +2\lambda Z^{2}\right)  X. \label{sd.11}%
\end{align}

At the infinity, it holds $1-X^{2}-Z^{2}=0$, thus the stationary points
$Q\left(  Z,X\right)  $ at the infinity are
\begin{equation}
Q_{1}=\left(  0,1\right)  ~\text{and }Q_{2}=\left(  0,-1\right)  .
\end{equation}
Hence, there are not any anisotropic solutions at the infinity value of
$\lambda$

However, $\lambda=3$, there exists the stationary point%
\begin{equation}
D^{\pm}=\left(  \pm\sqrt{1-Z^{2}},Z\right)
\end{equation}
where the asymptotic solution is anisotropic. That is a very interesting point
because $\lambda=3$ is the case where the future attractor in the finite
regime describes the de Sitter Universe.

On the surface $1-X^{2}-Z^{2}=0$, the effective equation of state parameter is
expressed as $w_{eff}\left(  X\right)  =-1+\frac{2}{3}\lambda X$. Thus, the
asymptotic solutions at the infinity are always scaling solutions, which
describe acceleration when $\lambda<1$ for point $Q_{1}$ or $\lambda>-1$ for
point $Q_{2}$.

The eigenvalues of the linearized system at the stationary points are
calculated $e_{1}\left(  Q_{1}\right)  =0$, $e_{2}\left(  Q_{1}\right)
=-2\lambda$ and $e_{1}\left(  Q_{2}\right)  =0$, $e_{2}\left(  Q_{2}\right)
=2\lambda$.

From the phase-space portraits presented in Fig. \ref{bf2}, we observe that
for $\lambda<0$ the stationary point $Q_{2}$ is a saddle point, and $Q_{1}$ is
a source, while for $\lambda>0$, $Q_{1}$ is a saddle point and $Q_{2}$ is a
source. For $\lambda<6$, the unique attractor is point $P_{2}$, while the
trajectories have origin the infinity and the surface of stationary points
$P_{1}$. For $\lambda>6$, we observe that the trajectories start from the
infinity, $Q_{2}$, and the source $P_{2}$, and they reach the surface of
points $P_{1}$ and the saddle point $Q_{1}$. Then through these lines, the
trajectories end at the infinity.

The latter means that for $\lambda>6$, the Universe moves from isotropy to
anisotropy and vice versa. While for $\lambda<6$, the final stage of the
Universe is the isotropy. Hence, in order $f\left(  T,B\right)  =T+F\left(
B\right)  $ to solve the isotropic problem $\lambda<6$.

Finally, for $\lambda=3$, from Fig. \ref{bf2} we remark that the family of
points $D^{\pm}$ are saddle points, and in this case, the future attractor is
the de Sitter Universe.

\section{Integrability and analytic behaviour}

\label{sec5}

We proceed with our analysis with the investigation of the integrability
properties for the two dynamical systems (\ref{sd.06}), (\ref{sd.07}). This
kind of analysis is essential because we can relate the numerical behaviour of
the trajectories presented in Figs. \ref{bf1} and \ref{bf2} to actual
solutions of the dynamical system.

An equivalent way to write the dynamical system is to the second-order
ordinary differential equation
\begin{equation}
\Sigma\left(  \frac{d^{2}\Sigma}{d\tau^{2}}+\left(  \lambda-6\right)
\frac{d\Sigma}{d\tau}\right)  -2\left(  \frac{d\Sigma}{d\tau}\right)
^{2}-\lambda\Sigma^{3}\left(  \frac{d\Sigma}{d\tau}\right)  =0. \label{da.00}%
\end{equation}

We perform the change of variable $\Sigma=Y^{-1}$, thus the latter equation
becomes%
\begin{equation}
\frac{d^{2}Y}{d\tau^{2}}+\left(  \left(  \lambda-6\right)  -\frac{\lambda
}{Y^{2}}\right)  \frac{dY}{d\tau}=0,
\end{equation}
that is,%
\begin{equation}
\frac{dY}{d\tau}+\left(  \left(  \lambda-6\right)  Y+\frac{\lambda}{Y}\right)
=Y_{0}, \label{da.01}%
\end{equation}
where the latter can be solved by quadrature.

For large values of $Y$, the term $\left(  \lambda-6\right)  Y$ dominates in
equation (\ref{da.01}) such that the analytic solution to be approximated as%
\begin{equation}
Y\left(  \tau\right)  \simeq Y_{1}e^{-\left(  \lambda-6\right)  \tau},
\end{equation}
which means that for $\lambda<6$ when $\tau>0$, $Y\left(  \tau\right)  $
increases and $\Sigma\left(  \tau\right)  $ becomes zero. That describes the
behaviour near the isotropic universe. On the other hand for $\lambda>6$, the
solution is going far from the isotropic solution.

Furthermore, for small values of $Y\left(  \tau\right)  $, it follows
\begin{equation}
Y\left(  \tau\right)  \simeq Y_{1}+Y_{2}\tau\label{da.02}%
\end{equation}
which holds only for small values of $\tau$. The latter solution describes the
evolution of the trajectories near to the infinity.

\subsection{Painlev\'{e} analysis}

The Singularity analysis, which Sophie Kowalewski introduced
\cite{Kowalevski88} and established by the French school of Painlev\'{e}
\cite{pp1,pp2,pp3}, is a powerful method in order to make inferences about the
integrability properties of a given dynamical system and write the analytic
solution by using analytic functions. Indeed, a differential equation which
possesses the Painlev\'{e} property, its solution is written in terms of
Laurent expansions.

At this point, it is essential to clarify that in this section, the term
singularity, we mean movable singularities of the differential equations and
the term singularity should not be confused with the cosmological singularity
of the physical space.

The modern treatment of the singularity analysis is summarized in a simple
algorithm proposed by Ablowitz, Ramani and Segur (ARS algorithm)
\cite{ars1,ars2,ars3}. A pedagogical discussion on the ARS algorithm can be
found in the review \cite{buntis}.

Applications of the singularity analysis in cosmology cover various subjects,
from the study of anisotropic models \cite{CotsakisLeach, Demaret,pb3} to the
investigation of the integrability properties of dark energy theories
\cite{pb,pb2}.

Hence, the application of the ARS algorithm in equation (\ref{da.00}) gives
that the leading-order term is%
\begin{equation}
\Sigma\left(  \tau\right)  =\Sigma_{0}\tau^{-\frac{1}{2}}, ~-2\lambda\left(
\Sigma_{0}\right)  ^{2}=1
\end{equation}
with resonances
\begin{equation}
r_{1}=-1~\text{and }r_{2}=\frac{1}{2}\text{.}%
\end{equation}

The analytic solution of equation (\ref{da.00}) is presented in terms of Right
Puiseux Series, that is,
\begin{equation}
\Sigma\left(  \tau\right)  =\Sigma_{0}\tau^{-\frac{1}{2}}+\Sigma_{1}%
+\Sigma_{2}\tau^{\frac{1}{2}}+\Sigma_{3}\tau+... \label{da.03}%
\end{equation}
in which $\Sigma_{1}$ is arbitrary, the second integration constant, and
\[
\Sigma_{2}=-\frac{1+12\Sigma_{0}^{2}+3\Sigma_{1}^{2}}{4\Sigma_{0}},~\Sigma
_{3}=-\frac{\Sigma_{1}\left(  1+12\Sigma_{0}^{2}-2\Sigma_{1}^{3}\right)
}{5\Sigma_{0}^{2}},etc.\text{. }%
\]

Solution (\ref{da.03}) for small values of $\tau$ indicates that
$\Sigma\left(  \tau\right)  \rightarrow\infty$, which means that the
leading-order behaviour describes an anisotropic universe at the infinite
regime of the original dynamical system.

Furthermore, we observe that the analytic solution is expressed by a Right
Painlev\'{e} expansion, from the latter, we can infer that the leading-order
behaviour describes an unstable solution \cite{pb3}. That is in agreement with
the analysis presented before.

\section{Conclusions}

\label{sec6}

The dynamical system analysis for the field equations for the anisotropic
Bianchi I spacetime in modified teleparallel $f\left(  T,B\right)  $-theory
was performed. We determined the stationary points for the field equations in
dimensionless variables, in the so-called $H$-normalization, in the finite and
infinity regimes. Moreover, we used a Lagrange multiplier and introduced a
scalar field to attribute the higher-order degrees of freedom provided by the
theory.\ Thus a scalar field potential $V\left(  \phi\right)  $ has been
introduced. For our analysis we considered the exponential potential function
$V\left(  \phi\right)  =V_{0}e^{-\lambda\phi}$.

In the finite regime, we derived two families of stationary points, $P_{1}$
and $P_{2}$. Points $P_{1}$ describe a family of anisotropic Bianchi I
spacetimes. The spacetimes do not describe acceleration, while the asymptotic
solutions are always unstable. The points are sources or saddle points. On the
other hand, point $P_{2}$ corresponds to a spatially flat FLRW geometry which
describes acceleration for $\lambda<4$. $P_{2}$ is an attractor for values of
$\lambda\,<6$. We remark that for $\lambda=3$, the de Sitter Universe is recovered.

We introduced a new set of variables by defining a Poincare map and
investigating the evolution of the trajectories at the infinity. It was found
that for an arbitrary value of $\lambda$, two stationary points at the
infinity describe isotopic spatially flat FLRW geometries exist. However, for
the particular case in which $\lambda=3$, anisotropic asymptotic solutions
exist at the infinity. However, it was found that the stationary points at the
infinity always describe unstable solutions, and the points are sources or
saddle points.

According to the above analysis and with the numerical results presented in
Fig. \ref{bf2}, for the evolution of the trajectories for the field equations,
we remark that for $\lambda<6$, the trajectories have the origin at the
infinity \ or on the surface of anisotropic solutions described by $P_{1}$,
and the final attractor is point $P_{2}$. Hence, in order to solve the
isotropic problem in $f\left(  T,B\right)  $-theory with initial conditions
that of Bianchi I geometry, the parameter $\lambda$ should be constrained as
$\lambda\,<6$.

Furthermore, we investigated the integrability properties of the field
equations by using the singularity analysis. Such analysis is essential
because we know that the numerical trajectories correspond to actual solutions
to the problem for an integrable dynamical system. In the context of the
$H$-normalization, the field equations can be written as one second-order
ordinary differential equation.

We applied the ARS algorithm, and we found that the second-order ordinary
differential equation admits the Painlev\'{e} property with leading-order term
$\tau^{-\frac{1}{2}}$ and resonances $r_{1}=-1~$and $r_{2}=\frac{1}{2}$.
Hence, the analytic solution is expressed by a Right Puiseux Series. Indeed,
the leading-order term describes an anisotropic solution on the surface of
points $P_{1}$. Because the solution is written in the form of a Right Puiseux
Series, we can conclude that the singular solution $\tau^{-\frac{1}{2}}$ is
unstable, as was found by the analysis of the stationary points.

\textbf{Data Availability Statements:} Data sharing not applicable to this
article as no datasets were generated or analyzed during the current study.


\begin{thebibliography}{999}                                                                                              %


\bibitem {st1}B.P. Abbot et al. Phys. Rev. Lett. 123, 011102 (2019)

\bibitem {st2}E. Berti et al. Class. Quantum Grav. 32, 243001 (2015)

\bibitem {st3}D. Ayzenberg and C. Bambi, Tests of General Relativity using
black hole X-ray data, Handbook of X-ray and Gamma-ray Astrophysics (Eds. C.
Bambi and A. Santangelo, Springer Singapore, expected in 2022) [arXiv:2111.13918]

\bibitem {Teg}M. Tegmark et al., Astrophys. J. 606, 702 (2004)

\bibitem {Kowal}M. Kowalski et al., Astrophys. J. 686, 749 (2008)

\bibitem {Komatsu}E. Komatsu et al., Astrophys. J. Suppl. Ser. 180, 330 (2009)

\bibitem {suzuki11}N. Suzuki et. al., Astrophys. J. 746, 85 (2012)

\bibitem {Ade15}Planck Collaboration: P.A.R. Ade et al., A\&A 594, A13 (2016)

\bibitem {ade18}Planck Collaboration: Y. Akrami et al. A\&A 641, A10 (2020)

\bibitem {hot}E. Di Valentino, O. Mena, S. Pan, L. Visinelli, W. Yang, A.
Melchiorri, D.F. Mota, A.G. Riess and J. Silk, Class. Quantum Grav. 38, 153001 (2021)

\bibitem {guth}A. Guth, Phys. Rev. D 23, 347 (1981)

\bibitem {nh1}G.W. Gibbons and S.W Hawking, Phys. Rev. D 15, 2738 (1977)

\bibitem {nh2}S.W.\ Hawking and J.G. Moss. Phys. Lett. B 110, 35 (1982)

\bibitem {f1}K. Sato, MNRAS 195, 467 (1981)

\bibitem {f2}J.D\ Barrow and A. Ottewill, J. Phys. A\ 16, 2757 (1983)

\bibitem {w1}R. Wald, Phys.\ Rev. D 28, 2118 (1983)

\bibitem {ba1}J.D. Barrow and J. Stein-Schabes, Phys.\ Lett. A 103, 6-7, 315 (1984)

\bibitem {sim1}K. Bolejko, Phys.\ Rev. \ D 97, 083515 (2018)

\bibitem {rv1}V. Sahni, Class. Quantum Grav. 19, 3435 (2002)

\bibitem {rv2}L.\ Perivolaropoulos, Six Puzzles for LCDM Cosmology, [arXiv:0811.4684]

\bibitem {clifton}T. Clifton, P.G. Ferreira, A. Padilla and C. Skordis, Phys.
Rept. 513, 1 (2012)

\bibitem {df1}S. Nojiri and S.D. Odintsov, IJGMMP 4, 115 (2007)

\bibitem {df2}S. Nojiri, S.D. Odintsov and V.K. Oikonomou, Phys. Rept. 692, 1 (2017)

\bibitem {Aref1}A.A. Starobinsky, Phys. Lett. B 91, 99 (1980)

\bibitem {r1}H.A. Buchdahl, Mon. Not. Roy. Astron. Soc. 150 1 (1970)

\bibitem {r2}T.P. Sotiriou and V. Faraoni Rev. Mod. Phys. 82 451 (2010)

\bibitem {r3}S. Nojiri and S.D. Odintsov, Phys. Rep. 505 59 (2011)

\bibitem {r4}A.A. Starobinsky, Phys. Lett. B 91 99 (1980)

\bibitem {r5}G. Cognola, E. Elizalde, S. Nojiri, S.D. Odintsov, L. Sebastiani
and S.\ Zerbini, Phys. Rev. D. 77 046009 (2007)

\bibitem {r6}D. Glavan and C. Lin, Phys. Rev. Lett. 124, 081301 (2020)

\bibitem {r7}M.V. de S. Silva and M.E. Rodrigues, Eur. Phys. J. C 78, 638 (2018)

\bibitem {r8}Y. Zhong, and D.S.C. Gomez, Symmetry 10, 170 (2018)

\bibitem {r9}T. Harko, F.S.N. Lobo, S. Nojiri and S.D. Odintsov, Phys. Rev.
D84, 024020 (2011)

\bibitem {r10}J. Wu, G. Li,\ T. Harko and S.-D. Liang, EPJC 78, 430 (2018)

\bibitem {fg15}S. Nojiri, S.D. Odintsov and V.K. Oikonomou, Phys. Rept. 692, 1 (2017)

\bibitem {fg16}S.D. Odintsov, V.K. Oikonomou and F.P. Fronimos, Nucl. Phys. B
958, 115135 (2020)

\bibitem {fg17}S.D. Odintsov and V.K. Oikonomou, Phys. Lett. B 805, 135437 (2020)

\bibitem {fg10}F.K. Anagnostopoulos, S. Basilakos and E.N. Saridakis, Phys.
Rev. D 100, 083517 (2019)

\bibitem {ein28}A. Einstein 1928, Sitz. Preuss. Akad. Wiss. p. 217; ibid p.
224, A.~Unzicker and T.~Case, physics/0503046

\bibitem {Hayashi79}K. Hayashi and T. Shirafuji, Phys. Rev. D 19, 3524 (1979);
Addendum-ibid. 24, 3312 (1982).

\bibitem {Maluf:1994ji}J.W.~Maluf, J.\ Math.\ Phys.\ 35 (1994) 335

\bibitem {md1}H.I.~Arcos and J.G.~Pereira, Int.\ J.\ Mod.\ Phys.\ D 13, 2193 (2004)

\bibitem {Weitzenb23}R. Weitzenb\"{o}ck, Invarianten Theorie, Nordhoff,
Groningen (1923)

\bibitem {Ferraro}G. Bengochea and R. Ferraro, Dark torsion as the cosmic
speed-up, Phys. Rev. D. 79, 124019 (2009)

\bibitem {ftSot}B. Li, T.P. Sotiriou and J.D. Barrow, Phys. Rev. D 83, 064035 (2011)

\bibitem {ftTam}N. Tamanini and C.G. Bohmer, Phys. Rev D 86, 044009 (2012)

\bibitem {st1a}S.H. Chen, J.B. Dent, S. Dutta and E.N. Saridakis, Phys.\ Rev.
D 83, 023508 (2011)

\bibitem {st2a}J.B. Dent, S. Dutta and E.N.\ Saridakis, JCAP 01, 009 (2011)

\bibitem {st3a}K. Bamba, C.-Q. Geng, C.-C. Lee and L.-W. Luo, JCAP 11, 021 (2011)

\bibitem {st4}W. El Hanafy and E.N. Saridakis, JCAP 09, 019 (2021)

\bibitem {st5}C. Xu, E.N. Saridakis and G. Leon, JCAP 07, 005 (2012)

\bibitem {st6}A. Paliathanasis, J.D, Barrow and P.G.L.\ Leach, Phys. Rev. D
94, 023525 (2016)

\bibitem {bh1}S. Bahamonte, C.G. Boehmer and M. Wight, Phys. Rev. D 92, 104042 (2015)

\bibitem {revtel}S. Bahamonte, K.F. Dialektopoulos, C. Escamilla-Rivera, V.
Gakis, M. Hendry, J.L. Said, J. Mifsud and E. Di Valentino, Teleparallel
Gravity: From Theory to Cosmology, [arXiv:2106.13793] (2021)

\bibitem {myr11}R. Myrzakulov, EPJC 72, 1 (2012)

\bibitem {ftb1}A. Paliathanasis, JCAP 1708, 027 (2017)

\bibitem {ftb2}L. Karpathopoulos, S.\ Basilakos, G. Leon, A.\ Paliathanasis
and M. Tsamparlis, Gen.\ Rel. Gravit. 50, 79 (2018)

\bibitem {ftb02}M. Caruana, G. Farrugia and J.L. Said, EPJC 80, 640 (2020)

\bibitem {ftb3}A. Paliathanasis, Phys.\ Rev. D 95, 064062 (2017)

\bibitem {ftb4}A.\ Paliathanasis and G. Leon, Eur. Phys. J. Plus 136, 1092 (2021)

\bibitem {ftb5}G.A. Rave-Franco, C. Escamilla-Rivera and J.L. Said, EPJC 80,
677 (2020)

\bibitem {ftb6}G.A. Rave-Franco, C. Escamilla-Rivera and J.L. Said, Phys. Rev.
D 103, 084017 (2021)

\bibitem {ftb7}A. Paliathanasis and G. Leon, f(T,B) gravity in a
Friedmann-Lema\^{\i}tre-Robertson-Walker universe with nonzero spatial
curvature, [arXiv:2201.12189] (2022)

\bibitem {ftb8}A. Paliathanasis, Universe 7, 150 (2021)

\bibitem {ftb9}S. N\'{a}jera, A. Aguilar, G.A. Rave-Franco, C.
Escamilla-Rivera and R.A. Sussman, Inhomogeneous solutions in f(T,B) gravity,
[arXiv:2201.06177] (2022)

\bibitem {ftb10}A. Paliathanasis, Kasner universes in $f(T,\hat{B})$ gravity (2022)

\bibitem {kasner1}E. Kasner, Am. J. Math. 43, 217 (1921)

\bibitem {kas1}K. Adhav, A. Nimkar, R. Holey, Int. J. Theor. Phys. 46, 2396 (2007)

\bibitem {kas2}S.M.M. Rasouli, M. Farhoudi and H.R. Sepangi, Class. Quantum
Grav. 28, 155004 (2011)

\bibitem {kas3}X.O. Camanho, N. Dadhich and A. Molina, Class. Quantum Grav.
32, 175016 (2015)

\bibitem {kas4}P. Halpern, Phys. Rev. D 63, 024009 (2001)

\bibitem {kas5}M.V. Battisti and G. Montani, Phys. Lett. B 681, 179 (2009)

\bibitem {zs}Y.B Zeldovich and A.A. Starobinsky, Sov. Phys. JETP 34, 1159 (1972)

\bibitem {bt}J.D. Barrow and M.S. Turner, Nature 291, 469 (1981)

\bibitem {HT}S.W. Hawking and R.J. Tayler, Nature 209, 1278 (1966)

\bibitem {Mis69}\ C.W. Misner, Astroph. J. 151, 431 (1968)

\bibitem {collins}C.B Collins and S.W. Hawking, Astroph. J. 180, 317 (1973)

\bibitem {JB1}J.D. Barrow, Mon. Not. R. astron. Soc. 175, 359 (1976)

\bibitem {WE}J. Wainwright and G. F. R. Ellis (editors), Dynamical Systems in
Cosmology, Cambridge University Press, (1997)

\bibitem {szek0}P. Szekeres, Commun. Math. Phys. 41, 55 (1975)

\bibitem {cop1}E.J. Copeland, A.R. Liddle and D.\ Wands, Phys. Rev. D 57, 4686 (1998)

\bibitem {mo1}J. Wainwright and G.F.R Ellis, Cambridge University Press,
Cambridge (1997)

\bibitem {mo2}A.A. Coley, Dynamical Systems and Cosmology, Springer, Dordrecht (2003)

\bibitem {mo3}L. Amendola and S. Tsujikawa, Dark Energy, Cambridge University
Press, Cambridge (2010)

\bibitem {mo5}G. Leon and A. Paliathanasis, EPJC 78, 753 (2018)

\bibitem {mo6}A. Cid, F. Izaurieta, G.\ Leon, P. Medina and D. Narbona, JCAP
1804, 041 (2018)

\bibitem {mo8}C.R. Fadragas and G. Leon, Class. Quantum Grav. 31, 195011 (2014)

\bibitem {mo9}M. Abdelwahab, R. Goswani and P.K.S. Dunsby, Phys.\ Rev. D 85,
083511 (2012)

\bibitem {cher1}P. Christodoulidis, D. Roest and E.I. Sfakianakis, JCAP 1911,
002 (2019)

\bibitem {and3}A.\ Paliathanasis and G. Leon, Class. Quantum Grav. 38, 075013 (2021)

\bibitem {mmf1}A.A. Coley, Phys.\ Rev. D 62, 023517 (2000)

\bibitem {mmf4}A.A. Coley and R.J. van den Hoogen, Phys.\ Rev. D 62, 023517 (2000)

\bibitem {col112}A. Coley and G. Leon, Gen.\ Rel. Grav. 51, 115 (2019)

\bibitem {col113}G. Leon, A. Coley and A. Paliathanasis, Annals Phys. 412,
168002 (2020)

\bibitem {col11}A.A. Coley, G. Leon, P. Sandin and J. Latta, JCAP 12, 010 (2015)

\bibitem {ae4}B. Alhulaimi, R.J. van den Hoogen and A. A. Coley, JCAP 17, 045 (2017)

\bibitem {ans1}R. J. van den Hoogen, A.A. Coley, B. Alhulaimi, S. Mohandas, E.
Knighton and S. O'Neil, JCAP 18, 017 (2018)

\bibitem {dn1}G. Leon, Class. Quantum Grav. 26, 035008 (2009)

\bibitem {paper2}G. Leon and A. Paliathanasis, Anisotropic Spacetimes in
$f(T,B)$ theory II: Kantowski-Sachs Universe

\bibitem {paper3}G. Leon and A. Paliathanasis, Anisotropic Spacetimes in
$f(T,B)$ theory III: LRS Bianchi III Universe

\bibitem {paper4}A. Paliathanasis, Anisotropic Spacetimes in $f(T,B)$ theory
I: Noether symmetry analysis

\bibitem {ns1}M. Tsamparlis and A. Paliathanasis, Symmetry 10, 233 (2018)

\bibitem {rfa}R. Ferraro and F. Fiorini, Phys. Rev. D 84, 083518 (2011)

\bibitem {dm1}C.-Q. Geng, C.-C. Lee, E.N. Saridakis and Y.-P. Wu, Phys. Lett.
B 704, 384 (2011)

\bibitem {dm2}M. Wright, Phys. Rev. D 93, 103002 (2016)

\bibitem {lan1}S. Capozziello, J. Matsumoto, S. Nojiri and S.D. Odintsov,
Phys. Lett B 693, 198 (2010)

\bibitem {lan2}A.N. Makarenko, IJGMMP 13, 1630006 (2016)

\bibitem {lan3}D. S\'{a}ez-G\'{o}mez, Phys. Rev. D 85, 023009 (2012)

\bibitem {Kowalevski88}S. Kowalevski, Acta Math 12 177 (1889)

\bibitem {pp1}P. Painlev\'{e}, Le\c{c}ons sur la th\'{e}orie analytique des
\'{e}quations diff\'{e}rentielles (Le\c{c}ons de Stockholm, 1895) (Hermann,
Paris, 1897). Reprinted, O$\!$euvres de Paul Painlev\'{e}, vol.~I,
\'{E}ditions du CNRS, Paris, (1973)

\bibitem {pp2}P. Painlev\'{e}, Bulletin of the Mathematical Society of France
28 201 (1900)

\bibitem {pp3}P. Painlev\'{e}, Acta Math 25 1 (1902)

\bibitem {ars1}M.J. Ablowitz, A. Ramani and H. Segur, \ Lettere al Nuovo
Cimento \textbf{23,} 333 (1978)

\bibitem {ars2}M.J. Ablowitz, A. Ramani and H. Segur, J. Math. Phys.
\textbf{21,} 715 (1980)

\bibitem {ars3}M.J. Ablowitz, A. Ramani and H. Segur, J. Math. Phys.
\textbf{21,} 1006 (1980)

\bibitem {buntis}A. Ramani, B. Grammaticos and T. Bountis, Physics Reports,
180, 159 (1989)

\bibitem {CotsakisLeach}S. Cotsakis and P.G.L. Leach, J. Phys. A: Math. Gen.
27 1625 (1994)

\bibitem {Demaret}J. Demaret and C. Scheen, J. Math. Phys. A: Math. Gen. 29 59 (1996)

\bibitem {pb3}A. Paliathanasis and P.G.L.\ Leach, Phys. Lett. A 381, 1277 (2017)

\bibitem {pb}J. Miritzis, P.G.L. Leach and S. Cotsakis, Grav. Cosmol, 6 282 (2000)

\bibitem {pb2}A. Paliathanasis, J.D. Barrow and P.G.L Leach, Phys. Rev. D 94,
023525 (2016)
\end{thebibliography}
\end{document}